\documentstyle[aps,preprint]{revtex}
\begin{document}
\draft
\title{Spin-reflection positivity of the Kondo lattice at half-filling}
\author{$^{1,2}$Takashi Yanagisawa and $^1$Yukihiro Shimoi}
\address{$^1$Fundamental Physics Section, Electrotechnical Laboratory\\
1-1-4, Umezono, Tsukuba, Ibaraki 305, Japan}
\address{$^2$Max Planck Institute for Physics of Complex Systems\\
Bayreuther Str.40, Haus 16, 01187 Dresden, Germany}
\maketitle
\begin{abstract}
We examine the spin-reflection positivity of the ground state of the Kondo
lattice model at half-filling with the antiferromagnetic and
ferromagnetic exchange couplings $J\ne0$.  For every
positive $U>0$, where U is the
Coulomb interaction between the conduction electrons, we can show that the
ground state is unique.
\end{abstract}
\vspace{7mm}
\pacs{75.30.Mb,75.20.Hr.}

\section{Introduction}

     Strongly-corrrelated electrons have been studied with considerable
effort. Their complete understanding is now still difficult.  Among the various
models the Kondo lattice model is important as a fundamental model for
heavy-fermion systems which are typical strongly-correlated-electron systems.
In strongly-correlated electrons, rigorous results are still rare although
they will provide us valuable information as bench marks.  Recently, exact
results were obtained in some limiting cases for the Kondo lattice.
\cite{sig91,sig92,yan94}
Recently, an idea of the spin-reflection positivity was introduced, proving its
validity for the strongly-correlated electrons at half-filling.
\cite{ken88,lie89}  This idea was
first succesfully applied to the Hubbard model for $U>0$ at half-filling and
$U<0$ at
every filling.\cite{lie89} Later it was shown that this method is valid for the
symmetric-periodic Anderson model.\cite{ued92}
The purpose of this paper is to investigate the spin-reflection positivity for
the Kondo lattice following the method in Ref.\cite{yan95}.  We show that
the ground state of the Kondo lattice ($J\neq 0$) has the
property of spin-reflection positivity at half-filling for $U>0$ where $U$ is
the Coulomb interaction between the conduction electrons.
In our method, the Coulomb
interactions between the conduction electrons are crucial in deriving an energy
inequality such as $E(C) \geq E(P)$ where $C$ is a coefficient matrix
of the eigenstates of Hamiltonian and $P$ is a semipositive definite
matrix defined by $P=(C^{\dag}C)^{1/2}$.
As we have pointed out first in Ref.\cite{yan95}, we can apply the Schwarz
inequality by using fermions in dealing with the local-spin operators, where we
investigated $J<0$ and $U>|J|/4$.  In this paper
we discuss this method in more details and show that it is straightforward to
generalize our method for any non-zero $J$ and $U>0$.

\newpage
\section{The spin-reflection property of the Kondo lattice}

\subsection{Antiferromagnetic Kondo lattice}

Let us consider the Kondo lattice model given as
\begin{equation}
H = -\sum_{\sigma,<i,j>} t_{ij}c^{\dag}_{i\sigma} c_{j\sigma}
-\frac{U}{2} \sum_{i\sigma}n_{ci\sigma} +
U \sum_{i} n_{ci\uparrow}n_{ci\downarrow} +
J \sum_{i} \sigma_i \cdot S_i ,
\end{equation}
where $\sigma_i$ and $S_i$ denote spin operators of the conduction electrons
and the localized spins, respectively. $c_{i\sigma}(c^{\dag}_{i\sigma})$ denote
annihilation (creation) operators of the conduction electrons and we write
$n_{ci\sigma}$=$c^{\dag}_{i\sigma}c_{i\sigma}$. The second term indicates the
chemical potential so that we consider the half-filling case.
What we will consider is the following statement.

$Proposition$ $A$
We assume that the lattice is bipartite. $<i,j>$ in eq.(1) denotes a pair of
sites where one is on the sublattice A and the other is on the B sublattice.
The number of the lattice
is finite and the lattice is connected which means that there is a connected
path of bonds between every pair of sites.  Then the ground state of the Kondo
lattice in eq.(1) for the antiferromagnetic-coupling $J>0$ and $U>0$ at
half-filling is unique.

$Remarks$ We show several remarks before going into a proof.  We have
introduced
the Coulomb interaction $U$ on each site
to show a uniqueness of the ground state.
We write the Kondo lattice model in the following form,
\begin{eqnarray}
H &=& -\sum_{<i,j>\sigma} t_{ij}c^{\dag}_{i\sigma} c_{j\sigma}
+ \sum_{i} [ \frac{1}{2}J_{\perp}(c^{\dag}_{i\uparrow} c_{i\downarrow}
 f^{\dag}_{i\downarrow} f_{i\uparrow} + c^{\dag}_{i\downarrow}c_{i\uparrow}
 f^{\dag}_{i\uparrow} f_{i\downarrow} ) + \frac{1}{4}J_z(n_{ci\uparrow}-
n_{ci\downarrow})(n_{fi\uparrow}-n_{fi\downarrow}) ]\nonumber\\
&+& U\sum_{i}n_{ci\uparrow}n_{ci\downarrow}
-\frac{U}{2}\sum_{i\sigma}n_{ci\sigma},
\end{eqnarray}
where $f_{i\sigma}$($f^{\dag}_{i\sigma}$) denote annihilation (creation)
operators of localized spins. $n_{ci\sigma}$ and $n_{fi\sigma}$ indicate the
number operators of
the conduction electrons and the localized spins, respectively.
We should work in the subspace where the condition
$n_{fi\uparrow}$+$n_{fi\downarrow}$=1 holds. In Ref.\cite{yan95} we
introduced the Lagrange multipliers in the Hamiltonian.
Of course, we do not
necessarily need to introduce the Lagrange multiplier to restrict the Hilbert
space.  This is only a matter of taste.
We have written the perpendicular- and z-component of exchange interaction
as $J_{\perp}$ and $J_z$, respectively.
Let us assume that $J=J_{\perp}=J_z$.
We make the electron-hole transformation for the up
spins: $c_{i\uparrow}\rightarrow c^{\dag}_{i\uparrow}$, $c^{\dag}_{i\uparrow}$
$\rightarrow$$c_{i\uparrow}$ for $i\in$$A$, $f_{i\uparrow}\rightarrow$
$-f^{\dag}_{i\uparrow}$, $f^{\dag}_{i\uparrow}\rightarrow$$-f_{i\uparrow}$
for $i\in$$A$ and  $c_{i\uparrow}$$\rightarrow$ $-c^{\dag}_{i\uparrow}$,
$c^{\dag}_{i\uparrow}$$\rightarrow$$-c_{i\uparrow}$ for $i$$\in$$B$,
$f_{i\uparrow}\rightarrow$ $f^{\dag}_{i\uparrow}$,$f^{\dag}_{i\uparrow}$
$\rightarrow$ $f_{i\uparrow}$ for $i$$\in$$B$ where we have assumed that
the lattice
is bipartite-divided into two disjoint sets $A$ and $B$. The spin-down
electrons
are unaltered, $c_{i\downarrow}$ $\rightarrow$ $c_{i\downarrow}$ and
$f_{i\downarrow}$ $\rightarrow$ f$_{i\downarrow}$.  In this transformation
the z-component of the total spin is invariant at half-filling: $S_z$=0
$\rightarrow$ $S_z$=0.
Then $H$ is transformed into
\begin{eqnarray}
\tilde{H}&=& -\sum_{<i,j>\sigma}t_{ij}c^{\dag}_{i\sigma}c_{j\sigma}
-U\sum_{i}n_{ci\uparrow}n_{ci\downarrow}+\frac{U}{2}\sum_{i\sigma}
n_{ci\sigma} + \sum_{i}[-\frac{1}{2}J_{\perp}(c_{i\uparrow}
c_{i\downarrow}f^{\dag}_{i\downarrow}f^{\dag}_{i\uparrow} +
c^{\dag}_{i\downarrow}c^{\dag}_{i\uparrow}f_{i\uparrow}
f_{i\downarrow})\nonumber\\
&+&\frac{1}{4}J_z(1-n_{ci\uparrow}-n_{ci\downarrow})(1-n_{fi\uparrow}-
n_{fi\downarrow})].
\end{eqnarray}
  We work in the $S_z$=0 subspace since $S^2$ and
$S_z$ are conserved and every energy eigenvalue has a corresponding
eigenfunction in this subspace. For $\tilde{H}$ the constraint should read
$n_{fi\uparrow}=n_{fi\downarrow}$.  Here let us comment on this constraint.
We set $Q_i \equiv n_{fi\uparrow}-n_{fi\downarrow}$.  It is easy to see that
$Q_i$ commutes with $\tilde{H}$ and $Q_j$ (for any $j$):
\begin{equation}
[Q_i,\tilde{H}]=0; [Q_i,Q_j]=0 (\forall i ,j).
\end{equation}
Therefore the total space is divided into disjoint subspaces which are
specified
by eigenvalues of $Q_i$.  The physical space is given by
$S_0 = \{ \psi (\neq 0)|Q_i\psi=0 (\forall i)\}$.  In this subspace, the wave
function satisfies
\begin{equation}
\tilde{H}\psi = E\psi,
\end{equation}
\begin{equation}
Q_i\psi = 0 (\forall i),
\end{equation}
which are basic equations in our discussion.

$Proof$  There are two kinds of electrons with spin up
and spin down.  Let $\psi^{\sigma}_{\alpha}$ be an orthonormal basis set
which is composed solely of spin-$\sigma$ c and f electrons.  We assume that
basis states are real.  We follow the method of Ref. \cite{lie89} and
the ground-state
wave function in the space $S_z$=0 is written as $\psi= \sum_{\alpha\beta}$
$C_{\alpha\beta}$$\psi^{\uparrow}_{\alpha}$$\otimes$$\psi^{\downarrow}_{\beta}$.
$C$=($C_{\alpha\beta}$) is called the coefficient matrix of $\psi$.
Now the expectation value
of $\tilde{H}$ is given by:
\begin{eqnarray}
F&=& Tr(C^{\dag}H^{\uparrow}_0C+CH^{\downarrow}_0C^{\dag})
- J_{\perp}\sum_{i}\frac{1}{2}Tr(M^{\uparrow}_{fci}CM^{\downarrow}_{cfi}
C^{\dag} + M^{\uparrow}_{cfi}CM^{\downarrow}_{fci}C^{\dag})\nonumber\\
&+&J_z\sum_{i}[\frac{1}{4}Tr(C^{\dag}N^{\uparrow}_{cfi}C
+ CN^{\downarrow}_{cfi}C^{\dag})
-\frac{1}{4}Tr(CN^{\downarrow}_{fi}C^{\dag} +
C^{\dag}N^{\uparrow}_{fi}C)-\frac{1}{4}Tr(CN^{\downarrow}_{ci}C^{\dag}
+C^{\dag}N^{\uparrow}_{ci}C)\nonumber\\
&+&\frac{1}{4}Tr(N^{\uparrow}_{ci}CN^{\downarrow}_{fi}C^{\dag}
+N^{\uparrow}_{fi}CN^{\downarrow}_{ci}C^{\dag})]\nonumber\\
&-&U\sum_{i}Tr(N^{\uparrow}_{ci}CN^{\downarrow}_{ci}C^{\dag})
+\frac{U}{2}\sum_{i}Tr(C^{\dag}N^{\uparrow}_{ci}C+CN^{\downarrow}_{ci}C^{\dag}).
\end{eqnarray}
The matrices are defined by the following,
\begin{mathletters}
\label{allequations} 
\begin{equation}
(H^{\sigma}_0)_{\alpha\alpha'} = <\psi^{\sigma}_{\alpha}|-\sum_{<i,j>}t_{ij}
c^{\dag}_{i\sigma}c_{j\sigma}|\psi^{\sigma}_{\alpha'}>,\label{equationa}
\end{equation}
\begin{equation}
(M^{\sigma}_{cfi})_{\alpha\alpha'} = <\psi^{\sigma}_{\alpha}|
c^{\dag}_{i\sigma}f_{i\sigma}|\psi^{\sigma}_{\alpha'}>,\label{equationb}
\end{equation}
\begin{equation}
(M^{\sigma}_{fci})_{\alpha\alpha'} = <\psi^{\sigma}_{\alpha}|
f^{\dag}_{i\sigma}c_{i\sigma}|\psi^{\sigma}_{\alpha'}>,\label{equationc}
\end{equation}
\begin{equation}
(N^{\sigma}_{cfi})_{\alpha\alpha'} = <\psi^{\sigma}_{\alpha}|
n_{ci\sigma}n_{fi\sigma}|\psi^{\sigma}_{\alpha'}>,\label{equationd}
\end{equation}
\begin{equation}
(N^{\sigma}_{ci})_{\alpha\alpha'} = <\psi^{\sigma}_{\alpha}|
n_{ci\sigma}|\psi^{\sigma}_{\alpha'}>,\label{equatione}
\end{equation}
\begin{equation}
(N^{\sigma}_{fi})_{\alpha\alpha'} = <\psi^{\sigma}_{\alpha}|
n_{fi\sigma}|\psi^{\sigma}_{\alpha'}>.\label{equationf}\\
\end{equation}
\end{mathletters}
Please note that these matrices are real ones.
{}From the definition, ($N^{\sigma}_{cfi})_{\alpha\alpha'}=\sum_{\beta}$
$<\psi^{\sigma}_{\alpha}|n_{ci\sigma}|\psi^{\sigma}_{\beta}>$
$<\psi^{\sigma}_{\beta}|n_{fi\sigma}|\psi^{\sigma}_{\alpha}>$
=($N^{\sigma}_{ci}N^{\sigma}_{fi})_{\alpha\alpha'}$.
We have the up-down symmetry: $N^{\sigma}_{ci}$=$N^{-\sigma}_{ci}$,
$N^{\sigma}_{fi}$=$N^{-\sigma}_{fi}$, $H^{\sigma}_0 = H^{-\sigma}_0$
and $M^{\sigma}_{fci}$=$M^{-\sigma}_{fci}$.  Variation of the functional F
with respect to $C$ leads to the following equation,
\begin{eqnarray}
EC&=&CH^{\downarrow}_0+H^{\uparrow}_0C-J_{\perp}\sum_{i}\frac{1}{2}(
M^{\uparrow}_{fci}C
M^{\downarrow}_{cfi}+M^{\uparrow}_{cfi}CM^{\downarrow}_{fci})
+J_z \sum_{i}[\frac{1}{4}(N^{\uparrow}_{cfi}C
+CN^{\downarrow}_{cfi})\nonumber\\
&-& \frac{1}{4}(CN^{\downarrow}_{fi}+N^{\uparrow}_{fi}C) - \frac{1}{4}(C
N^{\downarrow}_{ci}+N^{\uparrow}_{ci}C)
+\frac{1}{4}(N^{\uparrow}_{ci}CN^{\downarrow}_{fi}+N^{\uparrow}_{fi}C
N^{\downarrow}_{ci})]\nonumber\\
&-&U\sum_{i}
N^{\uparrow}_{ci}CN^{\downarrow}_{ci}
+ \frac{U}{2}\sum_{i}(CN^{\downarrow}_{ci}+N^{\uparrow}_{ci}C).
\end{eqnarray}
{}From the constraint equations $Q_i\psi=0$, $C$ must satisfy
\begin{equation}
N^{\uparrow}_{fi}C=CN^{\downarrow}_{fi}.
\end{equation}
We can easily show that this equation is equivalent to the constraint,
$n_{fi\uparrow}=n_{fi\downarrow}$ which indicates that we have no
singly-occupied f-electron sites.  From the equation in eq.(10), we obtain
$<n_{fi\uparrow}(1-n_{fi\downarrow})>=TrC^{\dag}N^{\uparrow}_{fi}C(1-N^{\downarrow}_{fi})=TrC^{\dag}N^{\uparrow}_{fi}(1-N^{\uparrow}_{fi})C=0$ because
$N^{\sigma}_{fi}$ is a diagonal matrix $diag(\sigma_1,\sigma_2,\cdots)$ where
the diagonal elements are 0 or 1: $\sigma_i=0$ or 1. Inversely, we set that
$<n_{fi\uparrow}(1-n_{fi\downarrow})>=0$.  Then
$0=TrC^{\dag}N^{\uparrow}_{fi}C(1-N^{\downarrow}_{fi})=TrC^{\dag}(N^{\uparrow}_{fi})^2C(1-N^{\downarrow}_{fi})^2=Tr(1-N^{\downarrow}_{fi})C^{\dag}N^{\uparrow}_{fi}N^{\uparrow}_{fi}C(1-N^{\downarrow}_{fi})=\parallel N^{\uparrow}_{fi}C(1-N^{\downarrow}_{fi})\parallel^2$, where the norm $||\cdot||$ is defined by $||A||^2=TrA^{\dag}A$.  This means that $N^{\uparrow}_{fi}C(1-N^{\downarrow}_{fi})=0$.
Similarly, we have $(1-N^{\uparrow}_{fi})CN^{\downarrow}_{fi}=0$.
Hence eq.(10) is followed.  More directly, we can show eq.(10) by calculating
$n_{fi\uparrow}\psi =
\sum_{\alpha\beta}C_{\alpha\beta}n_{fi\uparrow}\psi_{\alpha\beta} =
\sum_{\alpha\beta}\sum_{\alpha'\beta'}C_{\alpha\beta}|\psi_{\alpha'\beta'}><\psi_{\alpha'\beta'}|n_{fi\uparrow}|\psi_{\alpha\beta}>
=
\sum_{\alpha\beta\alpha'}C_{\alpha\beta}\psi_{\alpha'\beta}(N^{\uparrow}_{fi})_{\alpha'\alpha} = \sum_{\alpha\beta}(N^{\uparrow}_{fi}C)_{\alpha\beta}\psi_{\alpha\beta}$, where we denote the basis as
$\psi_{\alpha\beta}=\psi^{\uparrow}_{\alpha}\otimes\psi^{\downarrow}_{\beta}$.
We can obtain similarly
$n_{fi\downarrow}\psi =
\sum_{\alpha\beta}(CN^{\downarrow}_{fi})_{\alpha\beta}\psi_{\alpha\beta}$ and
eq.(10) is also followed.

Then we can obtain the energy $E(C)$ given by the right-hand side in eq.(7)
with two equations (9) and (10).
Now, the identity below is useful in the following discussion,\cite{com01}
\begin{eqnarray}
J_zTr(C^{\dag}N^{\uparrow}_{ci}CN^{\downarrow}_{fi}&+&C^{\dag}N^{\uparrow}_{fi}CN^{\downarrow}_{ci})\nonumber\\
&=& -J_z\frac{1}{z}TrC^{\dag}(zN^{\uparrow}_{ci}
-N^{\uparrow}_{fi})C(zN^{\downarrow}_{ci}
-N^{\downarrow}_{fi})\nonumber\\
&+& zJ_zTrC^{\dag}N^{\uparrow}_{ci}CN^{\downarrow}_{ci}
+\frac{1}{z}J_zTrC^{\dag}N^{\uparrow}_{fi}CN^{\downarrow}_{fi}\nonumber\\
&=& -J_z\frac{1}{z}TrC^{\dag}(zN^{\uparrow}_{ci}
-N^{\uparrow}_{fi})C(zN^{\downarrow}_{ci}
-N^{\downarrow}_{fi})\nonumber\\
&+& zJ_zTrC^{\dag}N^{\uparrow}_{ci}CN^{\downarrow}_{ci}
+\frac{1}{2z}J_zTr(C^{\dag}N^{\uparrow}_{fi}C+CN^{\downarrow}_{fi}C^{\dag}),
\end{eqnarray}
where z is a positive real number $z>0$ and we have used the relation in
eq.(10) to derive the second equality. Then the energy E(C) is written as
\begin{eqnarray}
E(C)&=& Tr(C^{\dag}H^{\uparrow}_0C+CH^{\downarrow}_0C^{\dag})
- J_{\perp}\sum_{i}\frac{1}{2}Tr(M^{\uparrow}_{fci}CM^{\downarrow}_{
cfi}
C^{\dag} + M^{\uparrow}_{cfi}CM^{\downarrow}_{fci}C^{\dag})\nonumber\\
&+&J_z\sum_{i}[\frac{1}{4}Tr(C^{\dag}N^{\uparrow}_{cfi}C
+ CN^{\downarrow}_{cfi}C^{\dag})
-\frac{1}{4}Tr(CN^{\downarrow}_{fi}C^{\dag} +
C^{\dag}N^{\uparrow}_{fi}C)-\frac{1}{4}Tr(CN^{\downarrow}_{ci}C^{\dag}
+C^{\dag}N^{\uparrow}_{ci}C)]\nonumber\\
&+&\sum_{i}[
-\frac{1}{4z}|J_z|TrC^{\dag}(zN^{\uparrow}_{ci}-N^{\uparrow}_{fi})C
(zN^{\downarrow}_{ci}-N^{\downarrow}_{fi})\nonumber\\
&+& \frac{1}{4}z|J_z|TrC^{\dag}N^{\uparrow}_{ci}CN^{\downarrow}_{ci}
+\frac{1}{8z}|J_z|Tr(C^{\dag}N^{\uparrow}_{fi}C+CN^{\downarrow}_{fi}C^{\dag})],\nonumber\\
&-&U\sum_{i}TrC^{\dag}N^{\uparrow}_{ci}CN^{\downarrow}_{ci}
+\frac{U}{2}\sum_{i}Tr(C^{\dag}N^{\uparrow}_{ci}C+CN^{\downarrow}_{ci}C^{\dag}).
\end{eqnarray}
Since the energy $E(C)$ is symmetric with respect to the spin, we can set that
$C$ is hermitian: $C=C^{\dag}$.  It is also easy to see that $C$ and $C^{\dag}$
satisfy the same Schr\"{o}dinger equation.
There is a hermitian positive semidefinite matrix $P$ which satisfies
$CC^{\dag}=P^2$, where $P$ is determined uniquely.\cite{lan72}  According to
the Schwarz inequality for a square matrix $M$,
\begin{equation}
|TrCMC^{\dag}M^{\dag}|\leq TrPMPM^{\dag},
\end{equation}
we obtain an inequality $E(C)\geq E(P)$ for $J>0$ and $U>z|J_z|/4$.
Since $z$ is an arbitrary positive real
number, we can choose $z$ so that $U>z|J_z|/4$ holds for any positive
$U$.  Therefore we have $E(C)\geq E(P)$ for every $U>0$.  Since we have assumed
that $C$ is the coefficient matrix of the ground state, we obtain $E(C)=E(P)$.
This indicates that there is a state with $C=P$ or $C=-P$ among the ground
states.  Here we will show that the new matrix $P$ also satisfies the
constraint $n_{fi\uparrow}=n_{fi\downarrow}$, i.e. $N_iP=PN_i$ where we set
$N_i \equiv N^{\uparrow}_{fi}=N^{\downarrow}_{fi}$.
Due to the Schwarz inequality
$TrCN_iCN_i \leq TrPN_iPN_i$, we have
$0 \leq <n_{fi\uparrow}(1-n_{fi\downarrow})>_P \equiv TrPN_iP(1-N_i) =$
$TrPN_iP-TrPN_iPN_i \leq TrCN_iC-TrCN_iCN_i = TrCN_iC(1-N_i) = 0$.
Then $TrPN_iP(1-N_i)=0$ is followed, which indicates that
 $TrPN_iP(1-N_i) = TrPN^2_iP(1-N_i)^2 = Tr(1-N_i)PN_iN_iP(1-N_i) =$
$\parallel N_iP(1-N_i)\parallel^2=0$. Hence $N_iP(1-N_i)=0$, i.e.
$N_iP=N_iPN_i$ holds.
Similarly we have $PN_i=N_iPN_i$.  Therefore we have obtained the constraint
equation for P given by,
\begin{equation}
N^{\uparrow}_{fi}P=PN^{\downarrow}_{fi}.
\end{equation}
This result shows that the equality $E(C)=E(P)$ has its meaning.

Now we will show that the ground state is unique following the argument of
Ref. \cite{lie89}.  The Schr\"{o}dinger equation reads
\begin{eqnarray}
EC&=&CH^{\downarrow}_0+H^{\uparrow}_0C-J_{\perp}\sum_{i}\frac{1}{2}(M^{\uparrow}_{fci}C
M^{\downarrow}_{cfi}+M^{\uparrow}_{cfi}CM^{\downarrow}_{fci})
+J_z \sum_{i}[\frac{1}{4}(N^{\uparrow}_{cfi}C
+CN^{\downarrow}_{cfi})\nonumber\\
&-& \frac{1}{4}(CN^{\downarrow}_{fi}+N^{\uparrow}_{fi}C) - \frac{1}{4}(C
N^{\downarrow}_{ci}+N^{\uparrow}_{ci}C)]\nonumber\\
&-&
\frac{1}{4z}|J_z|\sum_{i}(zN^{\uparrow}_{ci}-N^{\uparrow}_{fi})C(zN^{\downarrow}_{ci}-N^{\downarrow}_{fi})\nonumber\\
&+&\frac{1}{8z}|J_z|\sum_{i}(N^{\uparrow}_{ci}C+CN^{\downarrow}_{fi})
-(U-\frac{z}{4}|J_z|)\sum_{i}N^{\uparrow}_{ci}CN^{\downarrow}_{ci}
+ \frac{U}{2}\sum_{i}(CN^{\downarrow}_{ci}+N^{\uparrow}_{ci}C).
\end{eqnarray}
Let $R=P-C$; then $R$ is positive semidefinite and satisfies eq.(14).  Let us
define
$K$ as a kernel of $R$, i.e.$K=\{v|Rv=0\}$.  $C$ and $P$ are diagonalized by a
unitary
matrix $U$: $C=U^{\dag}$diag($\sigma_1$,$\cdots$,$\sigma_r$)$U$ and
$P=U^{\dag}$
diag($|\sigma_1|$,$\cdots$,$|\sigma_r|$)$U$ where
$\sigma_1$,$\cdots$,$\sigma_r$
are eigenvalues of $C$.  At least there is one positive $\sigma_i$,
such
that $\sigma_i$=$|\sigma_i|$; otherwise we have $C=-P$.
Thus $R=P-C$ has at least
one zero eigenvalue, which indicates that there is a vector $v$ satisfying
$Rv$=0.
Then we obtain:
\begin{eqnarray}
0&=&RH^{\downarrow}_0v-J_{\perp}\sum_{i}\frac{1}{2}(M^{\uparrow}_{fci}R
M^{\downarrow}_{cfi}+M^{\uparrow}_{cfi}RM^{\downarrow}_{fci})v
+J_z\sum_{i}[\frac{1}{4}RN^{\downarrow}_{ci}N^{\downarrow}_{fi}
v\nonumber\\
&-&\frac{1}{4}RN^{\downarrow}_{fi}v-\frac{1}{4}RN^{\downarrow}_{ci}v]
-\frac{1}{4z}|J_z|\sum_{i}(zN^{\uparrow}_{ci}-N^{\uparrow}_
{fi})R(zN^{\downarrow}_{ci}-N^{\downarrow}_{fi})v\nonumber\\
&-&(U-z|J_z|/4)\sum_{i}N^{\uparrow}_{ci}RN^{\downarrow}_{ci}v
+\frac{1}{8z}|J_z|\sum_{i}RN^{\downarrow}_{fi}v
+\frac{U}{2}\sum_{i}RN^{\downarrow}_{ci}v.
\end{eqnarray}
Since $v^tR=0$, $\sum_{i}[J_{\perp}v^t(M^{\uparrow}_{fci}RM^{\downarrow}_{cfi}+
M^{\uparrow}_{cfi}RM^{\downarrow}_{fci})v+(1/2z)J_zv^t(zN^{\uparrow}_{ci}-
N^{\uparrow}_{fi})R(zN^{\downarrow}_{ci}-N^{\downarrow}_{fi})v
+2(U-zJ_z/4)v^tN^{\uparrow}_{ci}RN^{\downarrow}_{ci}v]=0$.
holds.  Because $R$ is positive semidefinite and
$N^{\sigma}_{ci}$=$N^{-\sigma}_{ci}$, $N^{\sigma}_{fi}$=$N^{-\sigma}_{fi}$
and $M^{\sigma}_{cfi}$=$M^{-\sigma}_{cfi}$,
we have $v^tM^{\uparrow}_{fci}RM^{\downarrow}_{cfi}v=$
$v^tM^{\uparrow}_{cfi}RM^{\downarrow}_{fci}v=$
$v^tN^{\uparrow}_{ci}RN^{\downarrow}_{ci}v$
=$v^tN^{\uparrow}_{fi}RN^{\downarrow}_{fi}v=0$ and
then $RM^{\downarrow}_{fci}v=RM^{\downarrow}_{cfi}v=RN^{\downarrow}_{fi}v=$
$RN^{\downarrow}_{ci}v=0$ is followed.  If we substitute
$N^{\downarrow}_{fi}v$ for $v$, we
obtain $RN^{\downarrow}_{ci}N^{\downarrow}_{fi}v=0$.  As a result, $RH_0v=0$
follows.  Now, by successive application of $H_0$, $M$ and $N$, we can
construct
all the basis states by virtue of the connectivity. Thus, every vector is in
$K$.  This proves the uniqueness of the lowest
energy state for $J>0$ and $U>0$ because we can easily reach a
contradiction if we assume that there are two ground states \cite{lie89}.
Since the energy-expectation
value is continuous with respect to parameters involved in the
Hamiltonian there is no level crossing with respect to $J$.($q.e.d.$)

In the large-$U$ limit, $H$ is mapped onto
a spin-1/2 antiferromagnetic Heisenberg model.  Then we can say that

$Corollary$  We assume the same conditions in the $Proposition$ $A$.
Then for the Kondo lattice with $J>0$ and $U>0$ at half-filling,
the ground state has $S=0$.

\subsection{Ferromagnetic Kondo lattice}
Let us turn to investigate the Kondo lattice model $H$ with the ferromagnetic
coupling $J=J_{\perp}=J_z<0$ for the half-filled band.
We again assume that $\Lambda$ is bipartite and we
make the electron-hole tranformation for the up
spins: $c_{i\uparrow}\rightarrow c^{\dag}_{i\uparrow}$, $c^{\dag}_{i\uparrow}$
$\rightarrow$$c_{i\uparrow}$ for $i\in$$A$, $f_{i\uparrow}\rightarrow$
$f^{\dag}_{i\uparrow}$, $f^{\dag}_{i\uparrow}\rightarrow$$f_{i\uparrow}$
for $i\in$$A$ and  $c_{i\uparrow}$$\rightarrow$ $-c^{\dag}_{i\uparrow}$,
$c^{\dag}_{i\uparrow}$$\rightarrow$$-c_{i\uparrow}$ for $i$$\in$$B$,
$f_{i\uparrow}\rightarrow -f^{\dag}_{i\uparrow}$,$f^{\dag}_{i\uparrow}$
 $\rightarrow -f_{i\uparrow}$ for $i\in B$ where we have assumed that
 the lattice $\Lambda$
is bipartite-divided into two disjoint sets $A$ and $B$. Note that the signs
in front of f-electron operators are different from those for the case $J>0$.
The spin-down electrons
are unaltered, $c_{i\downarrow}$ $\rightarrow$ $c_{i\downarrow}$ and
$f_{i\downarrow}$ $\rightarrow$ f$_{i\downarrow}$.  In this transformation
the
z-component of the total spin is invariant: $S_z$=0 $\rightarrow$ $S_z$=0.
Then
$H$ is transformed into
\begin{eqnarray}
\tilde{H}&=& -\sum_{<i,j>\sigma}t_{ij}c^{\dag}_{i\sigma}c_{j\sigma}
-U\sum_{i}n_{ci\uparrow}n_{ci\downarrow}+\frac{U}{2}\sum_{i\sigma}
n_{ci\sigma} + \sum_{i}[\frac{1}{2}J_{\perp}(c_{i\uparrow}
c_{i\downarrow}f^{\dag}_{i\downarrow}f^{\dag}_{i\uparrow} +
c^{\dag}_{i\downarrow}c^{\dag}_{i\uparrow}f_{i\uparrow}
f_{i\downarrow})\nonumber\\
&+&\frac{1}{4}J_z(1-n_{ci\uparrow}-n_{ci\downarrow})(1-n_{fi\uparrow}-
n_{fi\downarrow})].
\end{eqnarray}
Clearly we can apply the method in the previous section and then we obtain
the inequality $E(C)\geq E(P)$ for $J<0$ and $U>0$. A similar identity to
eq.(11) is easy to derive for $J<0$:
\begin{eqnarray}
J_zTr(C^{\dag}N^{\uparrow}_{ci}CN^{\downarrow}_{fi}&+&C^{\dag}N^{\uparrow}_{fi}CN^{\downarrow}_{ci})\nonumber\\
&=& -|J_z|\frac{1}{z}TrC^{\dag}(zN^{\uparrow}_{ci}
+ N^{\uparrow}_{fi})C(zN^{\downarrow}_{ci}
+ N^{\downarrow}_{fi})\nonumber\\
&+& z|J_z|TrC^{\dag}N^{\uparrow}_{ci}CN^{\downarrow}_{ci}
+ \frac{1}{z}|J_z|TrC^{\dag}N^{\uparrow}_{fi}CN^{\downarrow}_{fi}.
\end{eqnarray}
We can prove that $C=P$ (or $C=-P$) is a unique solution of $E(C)=E(P)$.
Thus we have shown that the lowest-energy state is unique. Therefore

$Proposition$ $B$
If we assume the same conditions mentioned in $Proposion$ $A$ for the
Hamiltonian in eq.(17), then the ground state at half-filling is unique for
$J<0$ and $U>0$.

$Remarks$ If we assume that the A and B sublattices have the same number of
lattice sites, then the ground state of the Kondo lattice has $S=0$ since
in the large-$|J|$ limit, $H$ is mapped onto the spin-1 Heisenberg model.
\cite{tsu92} In
general, we may be able to consider the lattices where the number of sites in
the A sublattice $|A|$ is greater than that of the B sublattice $|B|$.  In this
case, the ground state may have a high spin $S=|A|-|B|$, which is proved by the
Perron-Frobenius theorem.  For example, the 1D odd-site model
with the open boundary condition has $S=1$ ground state, while if we impose the
periodic boundary condition, the ground state has $S=0$ for small clusters
according to a diagonalization method.

\subsection{Spin-correlation functions}

Our theorem for the Kondo lattice model may have many implications. Let us
consider the spin-correlation functions given as
$S_{fc}(i)\equiv <S^+_i\sigma^-_i>$,
$S_{ff}(i,j)\equiv <S^+_iS^-_j>$ and $S_{cc}(i,j)\equiv
<\sigma^+_i\sigma^-_j>$.
The spin-reflection positivity
implies that these correlation functions have definite signs for every $J$
($\neq 0$).\cite{she94}  After making the electron-hole transformation for
$J>0$, $S_{fc}(i)$ is written as
\begin{equation}
S_{fc}(i)=
-<c^{\dag}_{i\uparrow}f_{i\uparrow}c^{\dag}_{i\downarrow}f_{i\downarrow}> =
-TrC^{\dag}M^{\uparrow}_{cf}CM^{\downarrow}_{fc} \leq 0.
\end{equation}
In a similar manner, it is easy to obtain
\begin{equation}
S_{ff}(i,j)\leq 0; i \in A, j \in B,
\end{equation}
\begin{equation}
S_{cc}(i,j)\leq 0; i \in A, j \in B,
\end{equation}
\begin{equation}
S_{ff}(i,j)\geq 0; i \in A, j \in A,
\end{equation}
\begin{equation}
S_{cc}(i,j)\geq 0; i \in A, j \in A.
\end{equation}
Thus antiferromagnetic orderings are found for nearest-neighbor spins and
for c and f electrons on each site. The RKKY interactions between localized
spins are oscillating functions.  Instead, for the ferromagnetic coupling
$J<0$, $S_{fc}(i)$ shows a ferromagnetic order,
\begin{equation}
S_{fc}(i)= TrC^{\dag}M^{\uparrow}_{cf}CM^{\downarrow}_{fc} \geq 0.
\end{equation}
Note that we have chosen the different signs for f electrons in the
electron-hole transformation for $J<0$.  $S_{ff}(i,j)$ and $S_{cc}(i,j)$ have
same structures as the case for $J>0$.

\newpage
\section{Discussion}
In this paper we have applied the method of spin-reflection positivity to the
Kondo lattice model by writing the exchange interaction
with fermion operators of localized electrons.
We have shown that the Kondo lattice with the non-zero exchange couplings
$J\neq 0$ and $U>0$ at half-filling has a unique grond state and the total
spin is 0 where we have assumed that the A and B sublattices have the same
number of lattice sites. Our theory depends on the Schwarz inequality to derive
the equation $E(C)=E(P)$ where $C$ is the coefficient matrix of the ground
state
and $P$ is the semipositive definite matrix given by $P=(C^{\dag}C)^{1/2}$.
It is important that the constraint equation $N_iC=CN_i$, which represents
$n_{fi\uparrow}=n_{fi\downarrow}$, is conserved for $P$:
$N_iP=PN_i$. This is a highly non-trivial result. Our results can be
generalized to more general models where the number of the f-electron sites
is less than that of the conduction electrons.  For example, the two-impurity
Kondo model has a unique ground state which is continuous with respect to $J>0$
and $J<0$ as far as $U>0$.  A characteristic structure
of the two-impurity problem may be observed as a sharp crossover between
the RKKY regime and the on-site Kondo regime.\cite{joe88,yan91,and93}
The spin-reflection positivity implies the antiferromagnetic orderings between
the f and conduction electrons within each site as well as the nearest-neighbor
antiferromagnetic RKKY interactioons for $J>0$.  The RKKY interaction shows an
ocillating behavior with a period which is precisely equal to the lattice
constant($\times 2$) for the half-filled conduction band.

{}From a technical point of view, the fact that $Q_i$ commutes with Hamiltonian
$\tilde{H}$ and $Q_j$ is important because an eigenfunction of $\tilde{H}$ is
also an
eigenfunction of $Q_i$.  The total space is divided into disjoint subspaces
according to eigenvalues of $Q_i$. Let us comment here about the
Lagrange-multiplier method in Ref.\cite{yan95}.  We define
$H_{eff}=\tilde{H}+\sum_{i}\lambda_iQ_i$. Then basic equations in each subspace
are written as
\begin{equation}
H_{eff}\psi = E\psi,
\end{equation}
and
\begin{equation}
Q_i\psi = q_i\psi (\forall i),
\end{equation}
where $q_i$ takes $0,-1$ and 1. The variational condtion for $F\equiv
<H_{eff}>=<\tilde{H}>+\sum_i\lambda_i<Q_i>$ reads
$\partial F/\partial \lambda_i = <\psi |Q_i| \psi>=0$ which indicates
$q_i=0(\forall i)$.  Therefore we obtain the same equations as eqs.(5) and (6).
The conditions $<Q_i>=0$ project out the physical subspace $S_0$.  If we start
from a state which does not belong to $S_0$, we cannot obtain a correct
solution
in a diagonalization since they have different (discrete) quantum numbers.


\begin{references}
\bibitem{com01}This is basically due to identities $2xy=-(x-y)^2+x^2+y^2$ or
$2xy=(x+y)^2-(x^2+y^2)$.  Then we shift $x$ to $xz(z>0)$.
\end{references}
\end{document}